\documentclass[onecolumn,amsmath,amssymb,aps,prd,showpacs,showkeys,nofootinbib]{revtex4}

\begin{document}


\author{V.S.Popov}
\author{M.A.Trusov}
\email{trusov@itep.ru}

\affiliation{ITEP, Moscow, Russia}

\title{Feynman operator calculus and singular quantum oscillator}

\date{June 4, 2008}



\begin{abstract}
New applications of Feynman disentangling method in quantum
mechanics are studied and the time-dependent singular oscillator
problem is solved in this approach. The important role of
representation group theory is discussed in this context.
\end{abstract}

\maketitle

The method of disentangling expressions, containing
non-commuting operators (FDM), suggested by Feynman in
\cite{Feynman}, gave an elegant solution of the harmonic
oscillator excitation under the arbitrary time-dependent external
force. In further developments the FDM was applied to some other
non-stationary quantum mechanical problems, see
\cite{Popov_magnet,Popov_oscillator} and \cite{Popov_UFN} for
review. And it was shown that the transition matrix elements
calculation became much simpler if the FDM was supplied by some
considerations from the representation theory of $SU(2)$ or
$SU(1,1)$ groups.

In this paper we apply the FDM to a non-quadratic system ---
singular oscillator with a variable frequency $\omega(t)$. We
obtain the self-contained analytic expressions for the transition
amplitudes between states with definite quantum numbers (at $t\to
\pm\infty$) and calculate the generating functions for transition
probabilities.

The problems considered in
\cite{Feynman,Popov_oscillator} can be generalized to a model of a
singular oscillator with variable frequency:
\begin{equation}
\begin{gathered}
\Hat H=\frac{1}{2}p^2+\frac{1}{2}\omega(t)^2x^2+\frac{g}{8x^2},\\
0<x<+\infty, \quad g=\text{const},\quad g>-1,\quad \hbar=m=1.
\end{gathered}
\label{1}
\end{equation}
The frequency $\omega(t)$ is an arbitrary real time function.  As
usual, we propose the boundary conditions:
\begin{equation*}
\omega(t)\to\omega_{\pm} \quad \text{at} \quad t\to\pm\infty
\end{equation*}
which allows one to define the final and initial states of the
oscillator.

It is well known that at a fixed $t$ the instantaneous spectrum of
the Hamiltonian (\ref{1}) is equidistant (see, e.g.,
\cite{Landau}):
\begin{equation}
E_n=2\omega(n-j),\qquad
j=-\frac{1}{2}-\frac{1}{4}\sqrt{1+g},\qquad n=0,1,2,\dots .
\label{2}
\end{equation}

We note further that the operators
\begin{equation}
J_1=\frac{1}{4}(px+xp),\quad
J_2=\frac{1}{4}\left(x^2-p^2\right)-\frac{g}{16x^2},\quad
J_0=\frac{1}{4}\left(x^2+p^2\right)+\frac{g}{16x^2} \label{3}
\end{equation}
satisfy the standard commutation relations of the  $su(1,1)$
algebra:
\begin{equation}
[J_1,J_2]=-iJ_0,\qquad [J_2,J_0]=iJ_1,\qquad [J_0,J_1]=iJ_2 ,
\label{4}
\end{equation}
and the Hamitonian (\ref{1}) is a linear combination of operators
$J_0$ and $J_2$:
\begin{equation}
H(t)=\left(\omega^2(t)+1\right)J_0+\left(\omega^2(t)-1\right)J_2.
\label{1a}
\end{equation}
The instantaneous eigenfunctions of the Hamiltonian (\ref{1})
realize the irreducible unitary infinite-dimensional
representation of the non-compact $su(1,1)$ algebra. The
corresponding Casimir operator (``angular momentum'' squared)
proves to be a constant and can be calculated directly:
\begin{equation}
\mathbf{J}^2=J_0^2-J_1^2-J_2^2=\frac{g-3}{16}=j(j+1) \label{5}
\end{equation}
so the weight of this representation is $j$.

For the simplest case $\omega_+=\omega_-=1$ the initial and the
final states are the eigenfunctions of the $J_0$ operator:
\begin{equation*}
J_0\psi_n=\lambda_n\psi_n,\qquad \lambda_n=n-j.
\end{equation*}
According to Ref. \cite{Popov_oscillator}, the transition
amplitude between initial $|m\rangle$ and final $|n\rangle$ states
can be expressed in terms of the generalized Wigner function for
the irreducible representation of the $su(1,1)$ algebra with
weight $j$:
\begin{equation}
w_{mn}=\left|f^{(j)}_{n-j,m-j}\right|^2,\qquad n,m=0,1,2,\dots .
\label{6}
\end{equation}
The latter can be obtained by an analytic continuation of a
standard Wigner function for the compact $su(2)$ algebra; the
details of this technique were described in the Appendix A in
\cite{Popov_oscillator}. In group theory such a method is known as
the ``Weyl unitary trick'' .

A generalization of this approach to the case of unequal initial
and final frequencies $\omega_+\ne \omega_-$ is quite obvious, as,
according to (\ref{1a}), the transformation from the initial state
basis to the final state one is simply a unitary rotation around
the axis 1, i.e., an element of the quasi-unitary group $SU(1,1)$.
Omitting intermediate calculations (all details can be found in
\cite{Trusov}), we derive the following expression for the
transition probabilities between states $|m,\omega_-\rangle$ and
$|n,\omega_+\rangle$ (see Eq. (13) from Ref.
\cite{Popov_oscillator}, for comparison):
\begin{equation}
w_{mn}=\frac{L!}{((L-S)!)^2S!}\frac{\Gamma(L-2j)}{\Gamma(S-2j)}\rho^{L-S}(1-\rho)^{-2j}
\left[{}_2F_1(-S,L-2j;L-S+1;\rho) \right]^2, \label{7}
\end{equation}
where $j$ is defined in (\ref{2}) and
\begin{equation*}
L=\max(m,n),\qquad S=\min(m,n),\qquad L-S=|m-n|.
\end{equation*}
The $\rho$ parameter, $0\le\rho<1$, has the same sense as in the
well-known problem of transitions in a regular oscillator (see,
e.g., \cite{Perelomov, Baz}) and can be calculated from the
classical oscillator equation of motion. The formula (\ref{7})
furnishes the ultimate answer to the problem given.

The Gauss hypergeometric function ${}_2F_1$ in (\ref{7}) has its
first argument being integer and negative (or zero), so it reduces
to the Jacobi polynomial. Making necessary transformations, we
obtain from (\ref{7}):
\begin{equation}
\begin{gathered}
w_{mn}=\frac{m!}{n!}\frac{\Gamma(n-2j)}{\Gamma(m-2j)}\rho^{n-m}(1-\rho)^{-2j}
\left[P_m^{(n-m,-2j-1)}(1-2\rho)\right]^2, \quad n\ge m,\\
w_{mn}=\frac{n!}{m!}\frac{\Gamma(m-2j)}{\Gamma(n-2j)}\rho^{m-n}(1-\rho)^{-2j}
\left[P_n^{(m-n,-2j-1)}(1-2\rho)\right]^2, \quad m\ge n,
\end{gathered}
\label{7a}
\end{equation}
which coincides with the standard quantum-mechanical result for
the transitions in the time-dependent singular oscillator obtained
by means of the Shroedinger equation solution (see \cite{Manko}
and references therein). Consider some special cases of
(\ref{7}),(\ref{7a}):

a) The excitation of the ground (vacuum) state: $m=0$,
\begin{equation}
w_{0n}=\frac{\Gamma(n-2j)}{n!\Gamma(-2j)}\rho^n(1-\rho)^{-2j}.
\label{8}
\end{equation}
In particular, for $g=-1$ we have
\begin{equation}
j=-\frac{1}{2},\qquad w_{0n}=\rho^{n}(1-\rho). \label{8a}
\end{equation}
This is the utmost value of the $g$ constant, as at $g<-1$ the
particle is collapsing into the potential centre \cite{Landau}, so
the Hamiltonian is not the self-conjugated operator any longer. It
is significant that the expression (\ref{8a}) corresponds to the
distribution of charged scalar boson pairs, created in vacuum with
a definite momentum $\mathbf{p}$ under the influence of a variable
homogeneous electric field \cite{6, 7}.

b) At $g=0$ (i.e. $j=-3/4$) the problem considered corresponds to
the regular oscillator with time-dependent frequency on the
positive half-axis; in fact, the formulas (\ref{7a}) for this case
should coincide for the well-known formulas for the transition
probabilities between \textit{odd} oscillator states via Legendre
functions; the latter can be found in \cite{Perelomov, Baz}, for
example. And this really takes place, as there exist some
relations between Legendre functions and Jacobi polynomials with
specific parameter values; see \cite{Trusov} for details.

The last point to discuss is the generating functions for the
transition probabilities of the singular oscillator. For the
excitation of the ground state (the most interesting case in
practice) it is easy to obtain:
\begin{equation}
G_0(z)=\sum\limits_{n=0}^{\infty}
w_{0n}z^n=\left(\frac{1-\rho}{1-\rho z}\right)^{-2j}, \quad
|z|<1/\rho \label{9a}
\end{equation}
while for the transitions from the first excited state ($m=1$)
\begin{equation}
G_1(z)=\sum\limits_{n=0}^{\infty}
w_{1n}z^n=\left(\frac{1-\rho}{1-\rho
z}\right)^{-2j+2}\left(-2j\rho\left(\frac{1-z}{1-\rho}\right)^2+z\right),
\quad |z|<1/\rho \label{9b}
\end{equation}
The expressions (\ref{9a}) and (\ref{9b}) can be expanded
analytically for all the complex plane.

The formulas (\ref{9a}) and (\ref{9b})  are quite convenient to
compute various operator mean-values over the transition
probability distributions. In particular, for the adiabatic
invariant $I=\langle H\rangle/2\omega$ one obtains
\begin{equation}
\frac{I_+}{I_-}=\frac{\langle n-j
\rangle}{m-j}=\frac{1+\rho}{1-\rho}, \qquad m=0,1 . \label{10}
\end{equation}
One can show \cite{Trusov} that this expression remains valid for
transitions from an arbitrary level.

To summarize, we note that in our paper the transition
probabilities of the singular oscillator have been calculated for
an arbitrary frequency $\omega(t)$. To solve the problem, the
modified FDM was applied and the representation theory for
non-compact $su(1,1)$ algebra has been used. The final result for
the transition amplitudes has been presented in a self-contained
form, which is rather convenient for further applications. For the
transitions from low-lying states the expressions for the
generating functions have been derived and the adiabatic invariant
variation at the oscillator evolution has been calculated.


\begin{acknowledgments}
This work was partially supported by the Russian Foundation for
Basic Research (grant No. 07-02-01116) and by the Ministry of
Science and Education of the Russian Federation (grant No. RNP
2.1.1. 1972). One of the authors (M.A.T.) also thanks for partial
support the President Grant No. NSh-4961.2008.2 and the President
Grant No. MK-2130.2008.2 .
\end{acknowledgments}

\end{document}